\documentstyle[psfig]{mn}
\title[$H_0$ and $q_0$ from the AGN linewidth distribution.]{Determining
the cosmological parameters from the linewidths of active galaxies.}
\author[C.M.Rudge \& D.J.Raine]{C.M.~Rudge$$ and D.J.~Raine$$
\\$$Astronomy Group, University of Leicester, University Road,
Leicester, LE1 7RH, UK.}
\begin{document}
\maketitle
\begin{abstract}
We have previously shown that the linewidth distribution in AGN can be
accounted for by an axisymmetric broad emission line region. In this
paper we show that the linewidth distribution changes with redshift
and that these changes are dependent on $H_0$ and $q_0$. We show that
relatively small samples of AGN at high redshift with measured
linewidth at half maximum can be used to distinguish between values of
$H_0$ and $q_0$. Furthermore larger low redshift samples can be used
to distinguish between luminosity functions and hence different models
of quasar evolution.
\end{abstract}

\section{Introduction}

It is generally expected that the width, $v$, of the broad emission
lines in active galactic nuclei (AGN) will show some dependence on the
luminosity, $L$, of the nucleus as it is reasonable to assume that in
higher luminosity systems the emitting gas have higher velocities. It
is known from observations that in variable sources linewidths vary in
this sense with luminosity, e.g.\ C{\sc iv} in NGC4151 (Fahey,
Michalitsianos \& Kazanas 1991\nocite{Fahey91}). However it is not
clear that the same depedence of linewidth upon luminosity can be
extended across samples of AGN. Observations of samples \cite{W93,P97}
show some evidence for a $v-L$ relation, with however a large
scatter. Linewidths must therefore be dependent also on another
parameter. It is known from observations that several properties of
AGN appear to be dependent upon the line of sight to the observer. In
radio loud systems the linewidth shows a clear dependence upon $R$
\cite{Wills86,Wills95}, the ratio of core to lobe radio power, which
is an indicator of viewing angle. While it is not clear that the broad
emission line regions (BLR) are the same in radio loud and radio quiet
systems it is not unreasonable to expect that the linewidths in radio
quiet systems will also be viewing angle dependent. While it is
possible that other parameters play a role, the minimal hypothesis is
that $v=v(L,i,\lambda )$ with the same functional form for all
systems, but different for the lines at various wavelengths $\lambda$
(since these have different widths). The viewing angle, $i$, is the
angle of the axis of the BLR to the line of sight, i.e.\ $i=0$ for
systems viewed face on. Since it is difficult to measure the angle to
the line of sight for individual objects with any confidence, this
hypothesis has to be tested statistically. To do so requires an
assumption about the angular distribution and the luminosity function
for AGN. For the angular dependence it is natural to expand in
multipole moments to as high an order as the data justifies.

In a previous paper \cite{RR98} we used a random distribution in
angle (cut off beyond some angle $i_{\ast}$), the luminosity function from
Boyle, Shanks \& Peterson \shortcite{Boyle88} and a dipolar dependence
on angle to construct a line width distribution function which could
be matched to observations. In that paper systems were generally at
low redshift and so cosmological effects were not considered. These
cosmological effects enter in two ways. First through the luminosity
function, and second through the minimum observable luminosity in the
sample at each redshift $z$. Comparing the linewidth distribution at
different values of redshift therefore provides, in principle, a
method to determine the cosmological parameters $H_{0}$ and
$q_{0}$. The aim of this paper is to test the feasibility of the
method in terms of the number of linewidth measurements required to determine
these parameters. In fact, since we do not have an observational
sample to test, we have to construct one; this requires assumptions
about the cosmology. We shall therefore consider the related question
of the number of linewidth measurements required to distinguish
between significantly different cosmologies, between  $H_{0}=50,\ 75\
{\mathrm and\ } 100$ and between $q_{0}=0.0$ and $q_{0}=0.5$.

The method depends on knowledge of the luminosity function which
itself depends on the cosmological parameters. This raises several
issues. First, whether the method is any better than using fits to the
luminosity function alone to determine the cosmology. While in
principle this would be possible, in practice the dependence is weak
and it is not easy to distinguish between different cosmological
models from the luminosity data alone \cite{Boyle94}. The implication
is that the dependence of the linewidth distribution on the cosmology
that arises through the luminosity function is also weak. However,
there is a strong dependence of the shape of the linewidth
distribution on the cut-off luminosity which is itself sensitive to
the cosmology. This has the following consequences for the way the
method is applied.

In principle we are seeking a solution for the cosmological parameters from
simultaneous fits to the AGN luminosity function and linewidth distribution.
However, the weak influence of the cosmology on the luminosity function
means that we can use an iterative approach. We assume values for $H_{0}$
and $q _{0}$ in the luminosity function $\Phi (L,z)$. We calibrate the
linewidth distribution at low redshift and then compute it at high redshift
for this $\Phi$. Fitting to the data (when it is available for sufficiently
many systems) gives a new $H_{0}$ and $q_{0}$, which we can use to
find a new $\Phi$. This procedure can be iterated to convergence. In fact,
the final effect of the weak dependence of the luminosity function on the
cosmology is that very little (if any) iteration is necessary.

A potential problem is that the luminosity function depends on intrinsic
evolution as well as the cosmology. However, we find that this shows up in
an identifiable manner in the linewidth distribution, so it can be allowed
for. Equivalently, we can use the method to solve simultaneously for the
cosmology and to constrain the AGN evolution. We conclude that to
distinguish between open and closed cosmologies (even if the Hubble
parameter is not taken as fixed from other observations) would require in
the range 100 to 500 line profiles at a resolution of 500 km s$^{-1}$
at a redshift of order 2. 

\section{Development of the \lowercase{$z$} dependent line width distribution}

\subsection{Line width distribution}

As in our previous paper \cite{RR98} we assume a model for the BLR in
which the FWHM is dependent primarily upon ionising luminosity and
viewing angle alone. While it is accepted that there are other
parameters which may have some effect on linewidth we have shown that
such a restriction still allows us to account for the linewidth
distribution. Furthermore, in an axisymmetric model of the BLR, it is
not unreasonable to expect these to be the most important physical
parameters affecting linewidth.  Thus the FWHM, $v$, of a given broad
emission line is taken as a function of the ionising luminosity and
the inclination the system. This function can be expanded in spherical
harmonics with luminosity dependent coefficients. In Rudge \& Raine
\shortcite{RR98} we showed that the distribution of linewidths in
low-redshift systems could be reproduced if this function were taken
to be axisymmetric and only the first two terms of this series were
retained and the coefficients taken to have a common dependence on
luminosity. The FWHM of a given emission line is then given by
\begin{equation}
v=(a+b\sin i)L_{44}^{\alpha }  \label{vil}
\end{equation}
where the constants $a$, $b$ and $\alpha $ are chosen for each emission
line, $i$ is the angle to the line of sight of the axis of the BLR and 
$L_{44}$ is the luminosity in units of 10$^{44}$ erg\, s$^{-1}$.

Since it is difficult to determine the line of sight angle in a given
system, at least with any accuracy, we are lead to consider the linewidth
distribution rather than the linewidths of individual objects. Assuming that
the inclination of AGN is random across the sky, the number of systems
per unit velocity range at each $v$ is given for objects at low redshift by 
\begin{equation}
N(v)=\int \frac{\sin i}{\left| \frac{{\mathrm d}v}{{\mathrm d}i}\right| }%
\Phi (L_{44}){\mathrm d}L_{44}  \label{lwd}
\end{equation}
where the luminosity function $\Phi (L_{44})$ gives the luminosity
distribution. To extend this to high redshift systems we need to take
account of the cosmological and intrinsic evolution in $\Phi$ and the
redshift dependence of the range of luminosity covered in flux-limited
surveys.

\subsection{Luminosity functions}

The shape of the predicted linewidth distribution will clearly have
some dependence on the shape of the luminosity function. It is known
from observations that the quasar population evolves with redshift
causing the luminosity function to change with redshift. Thus it is
natural to expect that the shape of the predicted linewidth
distribution curve must also change with redshift. There is still much
debate over how best to model this evolution. Once a model for the
evolution has been chosen, the parameters used to fit the data are
then also dependent on the choice of values of $H_{0}$ and
$q_{0}$. Thus we shall first assume a luminosity function and an
evolution model, and test for differences when $H_{0}$ and $q_{0}$ are
changed. We then test for differences when the luminosity function and
evolution model are changed.

The quasar population can evolve in a combination of two basic
ways. First, by pure luminosity evolution, where only the luminosity
of quasars evolves with redshift and the total number remains
constant. Second, by pure density evolution, where the total number
evolves with redshift. Current work appears to favour pure luminosity
evolution but it should be noted (e.g.\ Boyle et al.\
1994\nocite{Boyle94}) that there is still a great deal of
uncertainty and no obvious best choice model.

There are several models for the luminosity function available
including those of Boyle et al.\ \shortcite{Boyle94}, Pei
\shortcite{Pei95}, Maccacaro et al.\ \shortcite{Maccacaro91}, Hasinger
\shortcite{Hasinger98} and Boyle et al.\ \shortcite{Boyle98}. To begin
with we shall concentrate on that of Boyle et al.\ \shortcite{Boyle94}
because of the large sample size, the detailed information given for
different models of evolution and the provision of results for two
extreme values of $q_{0}$ (0.0 and 0.5) spanning the range of
possibilities. The luminosity functions of Hasinger
\shortcite{Hasinger98} and Boyle et al.\ \shortcite{Boyle98} should
provide an improvement on this, with Hasinger \shortcite{Hasinger98}
using higher quality ROSAT data and Boyle et al.\ \shortcite{Boyle98}
using ASCA data.  However neither of these analyses contain
information on the effect of changing $q_{0}$. The number of objects
in the ASCA sample is also small. More information should become
available for both of these luminosity functions in the near
future. Maccacaro et al.\ \shortcite{Maccacaro91} use only the
\emph{Einstein} EMSS (\emph{Extended Medium Sensitivity Survey},
Stocke et al.\ \shortcite{Stocke91}) objects as used by Boyle et al.\
\shortcite{Boyle94} and develop a similar broken power law model of
the luminosity function. Perhaps the best alternative to the
luminosity function of Boyle et al.\ \shortcite{Boyle94}, for our
purposes, is that of Pei \shortcite{Pei95}. This uses a combined
sample of around 1200 sources and tests two different models for the
luminosity function. Note however that this is an optical rather than
an X-ray luminosity function. While this may seem an advantage in the
following work, which focuses on the H$\beta$ optical emission line,
Maccacaro et al.\ \shortcite{Maccacaro91} emphasize the incompleteness
of optical surveys and question the methods used to correct for
this. In developing the redshift dependent linewidth distribution we
shall use the luminosity function of Boyle et al.\ \shortcite{Boyle94}
and use that of Pei \shortcite{Pei95} to test for differences in the
resulting distribution functions.

The luminosity function given in Boyle et al.\ \shortcite{Boyle94} gives
several possible fits to the observed data using different evolution models.
We have chosen to use models G and H of Boyle et al.\ \shortcite{Boyle94} as
they are the best pair of fits to the combined ROSAT and EMSS data using the
same evolution model but different values of $q_0$. The models are given in
the form of a double power law 
\begin{equation}
\begin{array}{lr}
\Phi(L)=\Phi^{\ast} L_{44}^{-\gamma_1} & L<L^{\ast} \\ 
\Phi(L)=\frac{\Phi^{\ast}}{L_{44}^{\ast(\gamma_1- \gamma_2)}}
L_{44}^{-\gamma_2} & L>L^{\ast}
\end{array}.
\label{lf}
\end{equation}

The dependence on $z$ is included by evolution of $L^{*}$. Luminosity
evolution rather than evolution of the space density is assumed for the
models G and H. Thus, the quasar luminosity evolves by scaling in the
following way: 
\begin{equation}
\begin{array}{lr}
L(z)=L(z=0)(1+z)^{k} & z<z_{max} \\ 
L(z)=L(z=0)(1+z_{max})^{k} & z>z_{max}.
\end{array}
\label{Lev}
\end{equation}

\begin{figure}
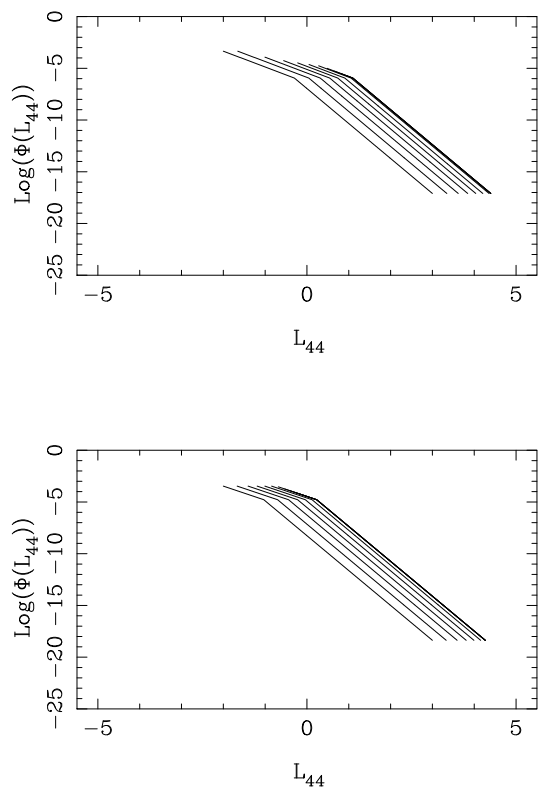

	\psfig{figure=lf0050.eps,angle=-90,width=\columnwidth}
	\psfig{figure=lf05100.eps,angle=-90,width=\columnwidth}
	\caption{Broken power law model of the luminosity function
	by Boyle et al.\ 1994. Both panels show
	$0<z<3$ with the plot moving to the right with increasing
	$z$. The upper panel is for $q_0=0,\ H_0=50$ and the lower
	panel is for $q_0=0.5,\ H_0=100$.}
	\label{fig:boylelf}
\end{figure}

It is important to note that (\ref{lf}) is in the form given in Boyle
et al.\ \shortcite{Boyle94} and is correct for the de-evolved
luminosity function i.e.\ the evolution model is applied to give all
systems as if they were at $z=0$. To introduce a $z$ dependence into
(\ref{lf}), i.e.\ to replace $\Phi(L)$ by $\Phi(L,z)$, the 2-power law
parameterization needs an extra scaling factor of $(1+z)^{\gamma_1 k}$
to ensure that $\Phi(L^{\ast},z)$ is constant over $z$. Note that
$\Phi^{\ast}$ is given in units of ${\mathrm 10^{-6}\, Mpc^{-3}\,
(10^{44}\, ergs\, s^{-1})^{-1}}$ and thus must be scaled by
$L^{\ast}_{44}(z=0)$ before entry into (\ref{lf}). The parameters for
the two models are given in table
\ref{tab:lf_fits}. Fig.\ \ref{fig:boylelf} shows how this luminosity
function evolves with $z$ and how the evolution changes with different
values of $q_0$ and $H_0$. Note for later reference that Boyle et al.\
\shortcite{Boyle93} suggested that the optical and X-ray luminosities
are related in the following way $L_{{\mathrm X}}\propto L_{{\mathrm
opt}}^{0.88\pm 0.08}$.

\begin{table}
\caption{Luminosity function parameters used for 2 power-law models G and H
of Boyle et al.\ \shortcite{Boyle94}.}
\label{tab:lf_fits}
\[
\begin{array}{ccccccc}
\vspace{5pt}
q_0 & \gamma_1 & \gamma_2 & \log L^{\ast}(z=0) & k & z_{max} & \Phi^{\ast}
\\
0.0 & 1.53 & 3.38 & 43.70 & 3.03 & 1.89 & 0.79 \\ \vspace{3pt}
0.5 & 1.36 & 3.37 & 43.57 & 2.90 & 1.73 & 1.45 \\
{\mathrm Errors} & \pm 0.15 & \pm 0.1 & \pm 0.2 & \pm 0.1 & \pm 0.1 &  \\ 
\end{array}
\]
\end{table}

\begin{figure}
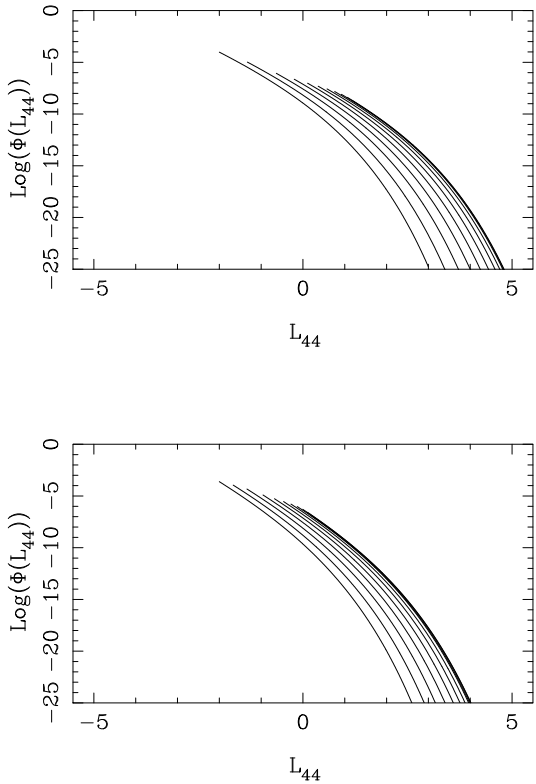

	\psfig{figure=peilf0150.eps,angle=-90,width=\columnwidth}
	\psfig{figure=peilf05100.eps,angle=-90,width=\columnwidth}
	\caption{Pei (1995) exponential model of the luminosity
	function. Both panels show $0<z<3$ with the plot moving to
	the right with increasing $z$. The upper panel is for
	$q_0=0.1,\ H_0=50$ and the lower panel is for $q_0=0.5,\
	H_0=100$.}
	\label{fig:peilf}
\end{figure}

The luminosity function of Pei \shortcite{Pei95} is given by
\begin{equation}
\Phi (L,z)=\frac{\Phi _{*}}{L_{z}}\left( \frac{L}{L_{z}}\right) ^{-\beta
}{\mathrm e}^{-\left( \frac{L}{L_{z}}\right) ^{1/4}}
\end{equation}
where the evolution model defines 
\begin{equation}
L_{z}=L_{*}(1+z)^{-(1+\alpha )}{\mathrm e}^{-(z-z_{*}^{2}/2\sigma _{*}^{2})}.
\end{equation}
The parameters for the model fits to the luminosity function are reproduced
in table \ref{tab:peilf}. Fig.\ \ref{fig:peilf} shows similar
information to fig.\ \ref{fig:boylelf}, but in this case for the
exponential fit rather than the broken power law fit of Boyle et al.\
\shortcite{Boyle94}.

\begin{table}
\caption{Parameters of the Exponential $L^{1/4}$ law luminosity function of
Pei \shortcite{Pei95}. Note that $h=H_0/100$.}
\label{tab:peilf}
\[
\begin{array}{lcc}
& (h,\ q_0,\ \alpha) & (h,\ q_0,\ \alpha) \\ \vspace{5pt}
{\mathrm Parameter} & (0.5,\ 0.5,\ -0.5) & (0.5,\ 0.1,\ -1.0) \\
\beta & 1.02\pm0.10 & 0.66\pm0.10 \\ 
z_{\ast} & 2.73\pm0.05 & 2.77\pm0.05 \\ 
\sigma_{\ast} & 0.93\pm0.03 & 0.91\pm0.03 \\ 
\log(L_{\ast}/L_{\odot} & 9.87\pm0.10 & 9.78\pm0.10 \\ 
\log(\Phi_{\ast}/Gpc^{-3}) & 5.33\pm0.17 & 4.34\pm0.20 \\
\end{array}
\]
\end{table}

We now have two models for the luminosity function at two different
values of $q_{0}$. These have both been produced under the assumption
that $H_{0}=50 {\mathrm km\,s^{-1}\,Mpc^{-3}}$. To use other values of
$H_{0}$ requires only an application of simple scaling factors to $L$
and $\Phi (L)$. These are
\begin{eqnarray}
L(H_{0} = 50) &=& L(H_{0})\left( \frac{H_{0}}{50}\right) ^{2}
\label{eqn:scale1} \\
\Phi (L,z,H_{0}) &=&\Phi (L,z,H_{0}=50)\left( \frac{H_{0}}{50}\right) ^{3}.
\label{eqn:scale2}
\end{eqnarray}
Equation (\ref{eqn:scale1}) scales the luminosity in a universe with
$H_{0} \neq 50$ to the value to be used in (\ref{lf}) where $H_{0}=50$.
Equation (\ref{eqn:scale2}) scales the calculated value of $\Phi (L,z)$ to
the value required for the universe where $H_{0}\neq 50$. These conversions
reflect the fact that a smaller value of $H_{0}$ would imply a more extended
universe, making luminosities greater but thinning out the galaxies, making
calculated densities less \cite{Weedman}.

\subsection{Limits of integration}

The redshift dependence of the linewidth distribution arises in just 2
places (assuming that the BLR itself does not change explicitly with
$z$).  The first is in the luminosity function as detailed above. The
second is through the $z$ dependence of the limits of integration in
(\ref{lwd}). This is partly accounted for in the intrinsic evolution
used in the luminosity function e.g.\  equation (\ref{Lev}). However
the lower limit of integration is also affected by the flux limit of
the observations. At some value of $z$ the faintest source we can see
is brighter than the evolved lower limit; in the Boyle et al.\ model
this occurs for $L_{min}^{observable}>L_{min}(1+z)^{k}$. So at any
value of $z$ the lower luminosity limit is the maximum of the two
minima (the observational limit and the evolved lower limit).

The minimum observable luminosity $L$ corresponding to a given flux
limit $F$ in the standard Freidmann cosmology (e.g.\ Marshall et al.\
1984\nocite{Marshall84}) is obtained from
\begin{equation}
F=\frac{L}{4\pi d_{L}^{2}},
\end{equation}
where the luminosity distance $d_{L}$ is given by 
\begin{equation}
\begin{array}{lclr}
d_{L} & = & \frac{c}{H_{0}q_{0}^{2}}[zq_{0}+(q_{0}-1)(\sqrt{1+2q_{0}z}-1)] & 
q_{0}>0 \\ 
d_{L} & = & \frac{cz}{H_{0}}\left[ 1+\left( \frac{z}{2}\right) \right]  & 
q_{0}=0
\end{array}
.
\end{equation}
Thus for a known flux limit and redshift it is possible to calculate the
luminosity distance, $d_{L}$, and hence the minimum observable luminosity at
this redshift. For the ROSAT observations used in Boyle et al.\
\shortcite{Boyle94}, $F\sim 4\times 10^{-15}{\mathrm ergs\,s^{-1}\,cm^{-2}}$.
In the following work, we assume that the samples generated will be
complete down to some flux limit. While any incompleteness,
particularly at the low luminosity end, will have some effect on the
shape of the linewidth distribution and consequently the application
to cosmology, for a sample of a few hundred objects at high redshift it
should be possible to generate a luminosity function specific to that sample.
It is likely that tailoring the luminosity function to the sample,
whether complete or not, will improve the accuracy of the results.

\subsection{New form for the linewidth distribution}

The linewidth distribution is now redefined to include the redshift
dependent luminosity function and integrated over $z$ so that the number of
systems, $N$, at each FWHM, $v$, is given by 
\begin{equation}
N(v) = \int \int_{L(z)_{min}}^{L(z)_{max}} \frac{\sin i}{\left |
\frac{{\mathrm d}v}{{\mathrm d}i} \right |} \Phi(L_{44},z) {\mathrm
d}L_{44}{\mathrm d}z.
\label{lwdz}
\end{equation}

\section{Testing}

We now develop a method to test the way in which the linewidth
distribution is affected by changes in $q_{0},\ H_{0}$ and the
luminosity function. Any significant differences in the shape of the
distribution curve will clearly become more obvious at higher redshift. Any
test therefore requires a sample of FWHM from high redshift AGN, say
$2.4<z<2.6$. As yet there are no published samples with a sizable number of
objects and FWHM measurements at high redshift. For example, the RIXOS
sample \cite{P97} has only 6 objects in the range $2<z<3$. The aim of the
current work therefore has to be to provide a case for carrying out
such observations by showing that the size of sample required is not
prohibitively large. This problem is approached in the following way.

First we produce a set of model parameters $a$, $b$ and $\alpha $ for
a sample of objects at low redshift where differing cosmology and
evolution models make little difference. This is done for both
luminosity functions. The value of $\chi ^{2}$ between the two model
distribution curves is minimized by choice of $a$ and $b$. Then, for
a chosen pair of values for $q_{0}$ and $H_{0}$ and luminosity
function, we use these parameters to produce a set of model
distribution curves at high redshift. Once observational data is
available this would then be compared with the theoretical distribution
curve, and $q_{0}$ and $H_{0}$ would be altered in an iterative process
until convergence was achieved. However, as this data is not
available, at this stage we generate a further model curve for
different values of the cosmological parameters. A small random sample
is then generated from the first distribution curve and a $\chi^{2}$
test carried out to find how large this sample needs to be so that
the second distribution curve differs from the first at the 95 per cent
confidence level.

We also consider whether the distribution curve is sensitive to the small
changes in the luminosity function model on changing $q_{0}$. This is
done by generating a distribution curve for say $q_{0}=0.0$ with the
luminosity function as for $q_{0}=0.5$ and comparing to the true case
where $q_{0}=0.0$ in the luminosity function. The fits of the
luminosity function to the observed data currently show no significant
improvement by changing $q_{0}$. We will show that the linewidth
distribution is more sensitive to the value of $q_{0}$ than is the
luminosity function and thus $q_{0}$ can be determined with a smaller
data set by this method.

This test of the effect of changing $H_0$ and $q_0$ is carried out on
the linewidth distribution produced using both luminosity functions. A
comparison between the two sets of distributions can then be carried
out to find which features of the distribution curve are most
sensitive to changes in the luminosity function. We can also test how
large a sample would be needed to distinguish between the luminosity
functions.

Note that for the $\chi ^{2}$ test the distribution curves are transformed
into histograms with bins of width 500\, km\, s$^{-1}$ to
reflect the expected resolution of observations. The random sample is
generated from this by a Monte Carlo type method to produce a similar
histogram. The resulting histograms are then normalized to contain the same
number of total objects. Thus the $\chi ^{2}$ test is not sensitive to the
absolute value of $N(v)$ only to the shape of the curve. The test is not
then sensitive to the evolution assumed in the luminosity function when only
a small range of $z$ is used. However it will be seen that plots of the peak
height of the distribution against $z$ do show very clear differences when
the evolution model is changed. Discarding evolution at this stage of
the testing is not unreasonable as the absolute value of $N(v)$ is
difficult to find observationally with a high degree of accuracy, since
this requires a high level of completeness in the sample. It will be
seen that the evolution model is testable separately by consideration
of the predicted dependence of $N(v)$ upon $z$.

\section{Results}

\subsection{Low redshift distribution}

\begin{figure}
	\psfig{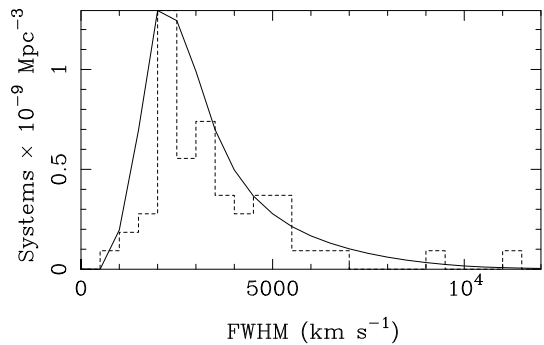}
	\caption{H$\beta$ linewidth distribution for the RIXOS
	objects with $z<0.5$.}
	\label{fig:hblowz}
\end{figure}

For a low redshift sample we have selected those objects in the RIXOS
sample  \cite{P97} with $z<0.5$ and measured FWHM H$\beta$. This
provides a sample of 54 objects including some for which the FWHM has
considerable uncertainty. Fig.\ \ref{fig:hblowz} shows the linewidth
distribution for these objects overlaid with the model curve for
$a=2500$\,km\,s$^{-1}$, $b=7000$\,km\,s$^{-1}$ and $\alpha=0.35$ using
the luminosity function of Boyle et al.\ \shortcite{Boyle94}. A
similar fit is produced with the Pei \shortcite{Pei95} luminosity
function using $a=1000,\ b=10000$ and $\alpha=0.31$. Note that
$i_{\ast}$, the angle beyond which the BLR is totally obscured to
view, is fixed at $60^{\circ}$ for all calculations.

\subsection{Evolution with $z$}

While the greatest differences in the distribution are naturally
expected to be seen at the highest redshifts, in order to make the
possibility of verification by observation realistic we have chosen
to use values for $z$ in the range $2.4<z<2.6$. This is separated from
$z=0$ by a large enough gap and covers a sufficient depth to provide a
reasonable chance of finding enough sources.  This range is also
towards the upper end of that covered by the luminosity function.

\begin{figure}
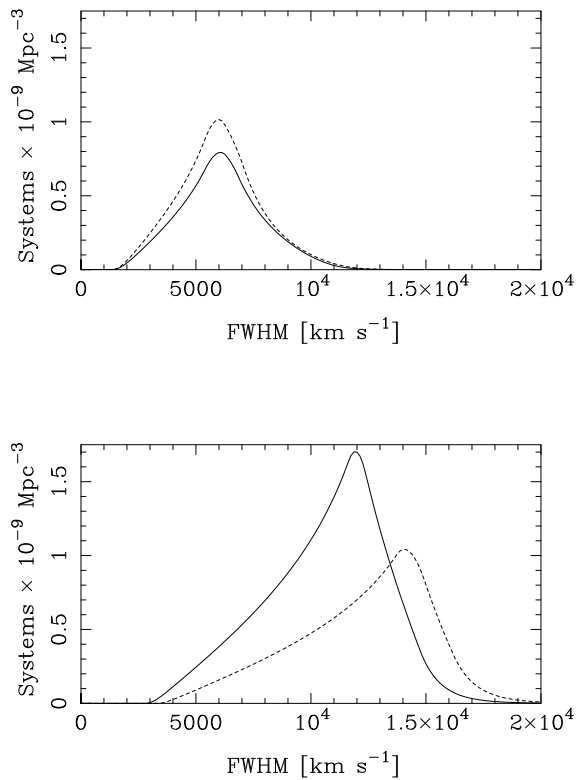

	\psfig{figure=cos1.eps,angle=-90,width=\columnwidth}
	\psfig{figure=cos2.eps,angle=-90,width=\columnwidth}
	\caption{Linewidth distribution at $0<z<0.5$ (upper) and
	$2.4<z<2.6$ (lower) for $q_0=0.5$ (solid) and $q_0=0.0$
	(dashed). All curves produced using the Boyle et al.\ 1994
	luminosity function.}
	\label{fig:lwdev}
\end{figure}
\begin{figure}
	\psfig{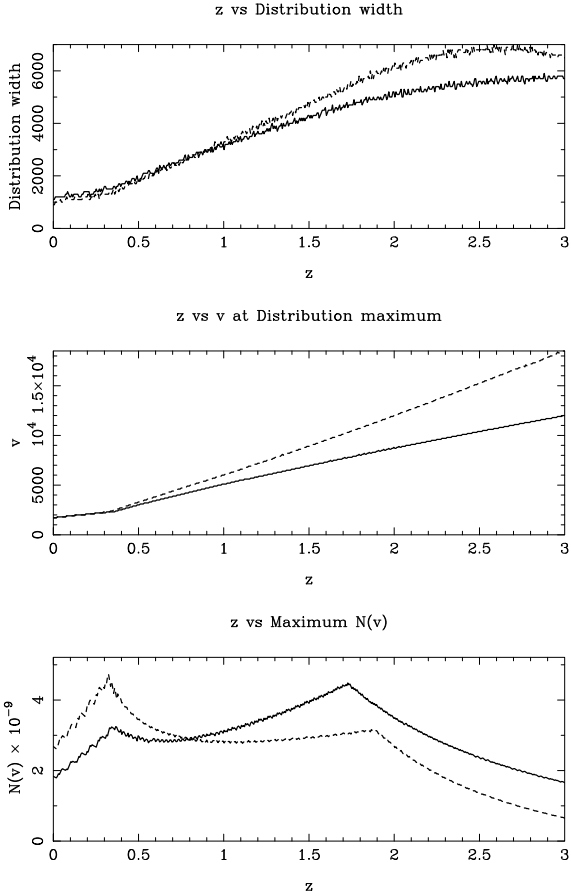}
	\caption{Change in the distribution of linewidths with $z$
	for $q_0=0.5$ (solid) and $q_0=0.0$ (dashed) produced using
	the Boyle et al.\ 1994 luminosity function. The upper panel
	shows the FWHM of the distribution, the middle panel shows
	the linewidth at maximum $N(v)$ and the lower panel shows
	the value of $N(v)$ at the peak.}
	\label{fig:params}
\end{figure}

Fig.\ \ref{fig:lwdev} shows how the distribution around $z=2.5$ changes
between $q_{0}=0.0$ and $q _{0}=0.5$. At high redshift there is a
clear difference between the two distributions with that for
$q_{0}=0.0$ having a peak at higher FWHM, but a smaller number of
systems at that value due to a broader distribution. However at low
redshift the distribution for $q_{0}=0.0$ has a narrower width
and hence a higher peak. Fig. \ref {fig:params} shows how three key
features of the shape of the distribution change with redshift.

\begin{enumerate}
\item  The full width at half maximum of the model curve.
\item  The linewidth at the peak value
\item  The value of $N(v)$ at this peak
\end{enumerate}

\begin{figure}
	\psfig{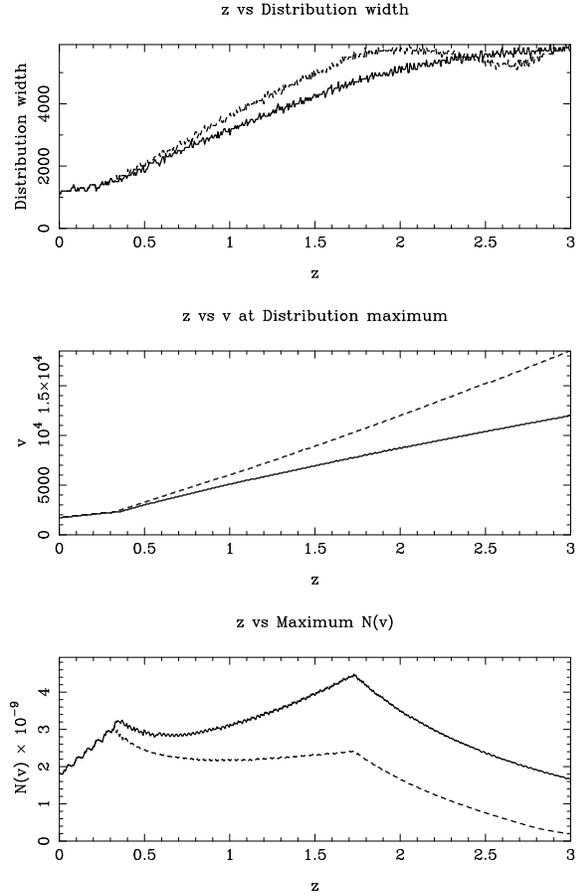}
	\caption{As for fig.\ \ref{fig:params} but with $q_0$ fixed
	at 0.5 in the luminosity function. Comparison with
	fig.\ \ref{fig:params} shows that the dependence of the
	distribution on $z$ is therefore not driven by changes in
	the luminosity function.}
	\label{fig:paramsfix}
\end{figure}

Fig.\ \ref{fig:paramsfix} shows the same information as
fig.\ \ref{fig:params} except that the shape of the luminosity function
is fixed as that for $q_0=0.5$ to test whether the changes in the
linewidth distribution with $q_0$ are caused primarily by the changes
in the luminosity function. Both figs. \ref {fig:params} and
\ref{fig:paramsfix} are produced using the Boyle et al.\
\shortcite{Boyle94} luminosity function.

\begin{figure}
	\psfig{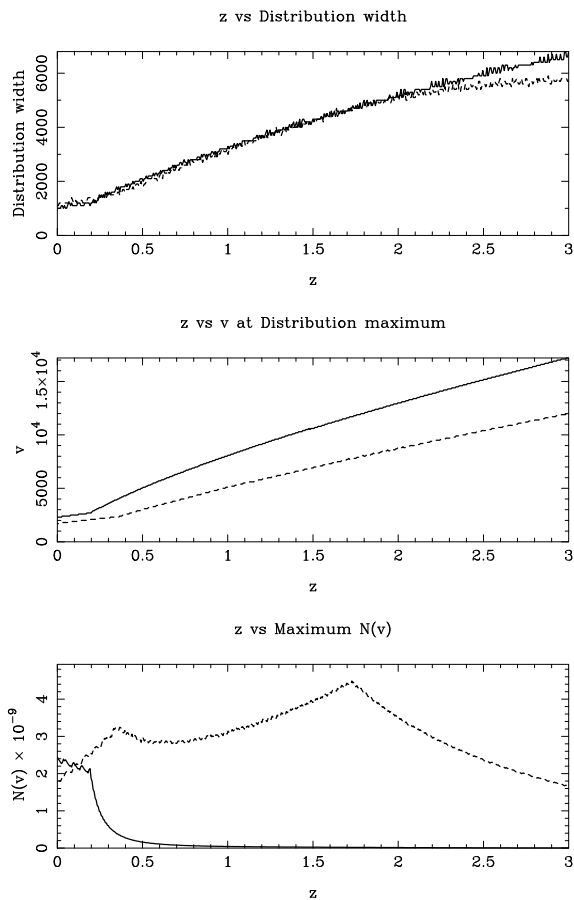}
	\caption{Comparison of changes in linewidth distribution
	with $z$ using the Pei (1995) luminosity function (solid)
	and the Boyle et al.\ (1994) luminosity function
	(dashed). Both distributions produced using $H_0=50$ and
	$q_0=0.5$. Individual panels are as for
	fig.\ \ref{fig:params}}
	\label{fig:params_comp}
\end{figure}

Fig.\ \ref{fig:params_comp} shows the change with redshift of the
distribution width, peak position and peak height produced with the
different luminosity functions, with $q _{0}=1.0$ and $H_{0}=50$. We
have carried out a similar $\chi ^{2}$ test on the distributions
produced at  $2.4<z<2.6$ to find how large a sample would be needed to
rule out one luminosity function assuming that the other is
correct. As the $\chi ^{2}$ test is not symmetrical the results are as
follows. If the Boyle luminosity function is correct then a sample of
about 150 objects would be needed to rule out that by Pei. Conversely,
if Pei has the correct function then about 250 objects would be
needed to rule out the Boyle form.

These first results show that changes in the value of $H_{0}$ and
$q_{0}$ do significantly affect the shape of the linewidth
distribution.  However we need the results of the $\chi ^{2}$ test to
show whether these differences can be observed with a reasonable size
sample. For each pair of $q_{0}$,\ $H_{0}$ values we have generated a
random sample of linewidths and tested this against other pairs of
values. Table \ref{tab:chisq} shows the minimum sample sizes to reject
the second pair of values in each case at the 95 per cent confidence level
using the Boyle et al.\ \shortcite{Boyle94} luminosity function.

\begin{table}
\caption{Minimum sample size of quasar linewidths for which the pair of $q_0$%
, $H_0$ values in the first column can be rejected when the sample is
generated using the pair of values in the first row of the table.
Distributions were generated using the Boyle et al.\ 1994 luminosity
function.}
\label{tab:chisq}
\[
\begin{array}{cc@{\hspace{20pt}}cccccc}
& q_0 & 0.0 & 0.0 & 0.0 & 0.5 & 0.5 & 0.5 \\ \vspace{5pt}
q_0 & H_0 & 50 & 75 & 100 & 50 & 75 & 100 \\
0.0 & 50 &  & 73 & 51 & 72 & 43 & 31 \\ 
0.0 & 75 & 142 &  & 106 & 769 & 82 & 51 \\ 
0.0 & 100 & 52 & 270 &  & 444 & 200 & 82 \\ 
0.5 & 50 & 121 & 5201 & 176 &  & 94 & 57 \\ 
0.5 & 75 & 52 & 142 & 1200 & 168 &  & 135 \\ 
0.5 & 100 & 34 & 65 & 165 & 71 & 447 &  \\
\end{array}
\]
\end{table}

Encouragingly many of the values in the table are of order 100 which is not
prohibitively large; hopefully samples of this size will exist in the
near future. However this is still beyond the scope of published
samples. Table \ref{tab:chisqpei} shows the same information as table \ref
{tab:chisq} except we use the Pei luminosity function and $q_0=0.1$ in
place of $q_0=0.0$.

\begin{table}
\caption{As for table \ref{tab:chisq} except in this case the distributions
were generated using the Pei (1995) luminosity function.}
\label{tab:chisqpei}
\[
\begin{array}{cc@{\hspace{20pt}}cccccc}
& q_0 & 0.1 & 0.1 & 0.1 & 0.5 & 0.5 & 0.5 \\ \vspace{5pt}
q_0 & H_0 & 50 & 75 & 100 & 50 & 75 & 100 \\
0.1 & 50 &  & 126 & 57 & 98 & 51 & 43 \\ 
0.1 & 75 & 201 &  & 176 & 3466 & 94 & 57 \\ 
0.1 & 100 & 71 & 407 &  & 407 & 400 & 98 \\ 
0.5 & 50 & 167 & 3199 & 176 &  & 94 & 57 \\ 
0.5 & 75 & 53 & 166 & 2332 & 166 &  & 131 \\ 
0.5 & 100 & 46 & 77 & 167 & 408 & 447 &  \\
\end{array}
\]
\end{table}

\section{Discussion}

These results show clearly that the linewidth distribution changes
with redshift. McIntosh et al.\ \shortcite{McIntosh98} find some
observational evidence for an increase in FWHM H$\beta$ between their
sample at $2<z<2.5$ and that of Boroson and Green \shortcite{BG92} at
low redshift, $z<0.5$. This dependence of FWHM upon $z$ may be biased
by selection of only higher luminosity objects in the high redshift
sample. More importantly this work shows that the evolution depends
significantly on the values of $H_{0}$ and $q_{0}$ and also on the
choice of luminosity function. The $\chi ^{2}$ test shows that we only
need samples of $\sim 100$ systems at $z=2.5$ to be able to
distinguish between extreme values of the cosmological parameters for
a given luminosity function.  While it is encouraging that these
sample sizes are not prohibitively large, the amount of FWHM data
required is still beyond that currently published. There is however
hope that the required data will be available in the near
future. Searches for clusters of quasars at high redshift, such as
those by Boyle et al.\ \shortcite{Boyle97} and Newman et al.\
\shortcite{Newman97}, may well provide the necessary information.

Perhaps the most significant result is that the major dependence on
$q_{0}$ is not within the luminosity function, as can be inferred
from fig.\ \ref{fig:paramsfix}. The X-ray luminosity function of Boyle et al.\
\shortcite{Boyle94} uses a large sample ($\sim 500$) of objects but
cannot distinguish successfully between values of $q _{0}$. Using a
linewidth distribution model should enable us to distinguish between
values of $q _{0}$ with samples of around 100 objects. Current
observational efforts to produce large samples of quasars (e.g.\ Boyle
et al.\ 1997\nocite {Boyle97}) will provide us with a good source of
candidates to observe further to obtain high resolution spectra from
which FWHM measurements can be made.

Comparison of luminosity functions is perhaps easier than testing for
$q_{0}$ and $H_{0}$ as initially only low redshift samples are needed.
These sample do need to be much larger ($\sim 500$) but there is much
more available data at low redshift. It may be that with a sample of
this size, it is not possible to get a good fit to the linewidth
distribution when certain luminosity functions are used. The
distribution when using the Pei \nocite{Pei95} function was noticeably
narrower than that for the Boyle \nocite{Boyle94} function. If this
does not prove possible, the large difference in the number of systems
at the distribution peak, across the full range of redshift, should be
a simple test of the evolution models with a small, but complete, high
redshift sample. Note that this statistic is not currently used in the
$\chi ^{2}$ testing due to the requirement of complete samples. In
fact the distribution curves are all normalized to have the same total
number of objects for the testing. Thus it is purely the shape of the
distribution that is tested and not the actual number of systems.

It should also be noted that current problems with the luminosity function
models include not only the accurate shape and evolution but also the
luminosity range. It has been suggested \cite{Hasinger98} that the AGN
luminosity function should extend to low luminosities and join the normal
galaxy luminosity function. Our results should be sensitive to this. 
Both of the luminosity functions used here are modelled with a specific low
luminosity cut-off. 

In this work, as in Rudge \& Raine \shortcite{RR98}, we have assumed
that the distribution of $\sin i$ is uniform and that the cut off
angle $i_{\ast}$ is constant. In reality this is probably not the
case. It is reasonable to expect that $i_{\ast}$ increases with
luminosity, i.e.\ the opening angle of AGN is greater in higher
luminosity systems. In Rudge \& Raine 1999 (unpublished) we have found
in the RIXOS sample \cite{P97} that this is in fact the case, while
the distribution of $\sin i$ is uniform in each luminosity bin. As a
result the distribution of $\sin i$ will be biased toward face-on
objects. While the effects that this has on the cosmological
predictions are not clear as yet, any such effect will be reduced by
removing the lower luminosity objects from samples. This, to some
extent, is done naturally at high redshift as we cannot observe the
lower luminosity sources assumed to be present. However reducing the
range of luminosity considered will probably also result in an
increase in the sample size needed.

With the exception of the recently developed supernovae observations,
all methods for obtaining $q_0$ suffer from uncertain evolutionary
effects which introduce a scatter comparable to the magnitude of the
effect being measured. Our method is no exception. What appears to be
surprising, however, is that not only can the data be used to model
the source evolution, but that this can be done in many cases with a
relatively modest number of observations.

While our work has concentrated on modelling the spread of linewidths
by an axisymmetric BLR other authors have suggested other parameters
than viewing angle to account for the scatter in the
linewidth--luminosity relation, e.g.\ $v=v(L,\alpha_{\rm X})$ model of
Wandel \& Boller \shortcite{Wandel98} and the model of Robinson
\shortcite{Robinson95} which has linewidth changing with profile
curvature.

In fact our results do not depend on the assumption made here that the
scatter in the $v$--$L$ relation is a viewing angle effect: any
analysis which attributes this scaater to a single additional
parameter will give similar results. Thus, while this method of
obtaining the cosmological parameters is unlikely ever to achieve the
accuracy or robustness of the supernova method, we believe it to be
worth pursuing as a viable supplementary approach.

\section{Conclusion}

In this paper we have shown that in an axisymmetric broad line region,
the linewidth distribution exhibits a strong dependence on $H_{0}$ and
$q _{0}$. More importantly we have shown that these dependencies
should be testable in the near future with observed samples of $\sim
100$ objects at high redshift. We have also shown that the changes are
not primarily a result of changes in the luminosity function due to
uncertainties in the value of $q_{0}$. It should also be possible to
distinguish between model luminosity functions using large, low
redshift samples or small, but complete, high redshift samples. This
provides the theoretical basis for a new method of determining the
values of $q_{0}$ and $H_{0}$ using observations of AGN. This
statistical method makes no attempt to find the distance and
luminosity of individual objects.

\section{acknowledgements}

CMR acknowledges the support of PPARC, in the form of a research
studentship. The authors wish to thank Gordon Stewart and Adam Blair for
discussion of the luminosity function.

\end{document}